\documentstyle[aps,prd,multicol]{revtex}
\tighten

\begin{document}
\title{Neutral heavy lepton production at next high energy $e^+e^-$ linear colliders}
\author{F.M.L. Almeida Jr.\thanks{E-mail: marroqui@if.ufrj.br},Y. A. 
Coutinho\thanks{E-mail: yara@if.ufrj.br}, \\
J. A. Martins Sim\~oes\thanks{E-mail: simoes@.if.ufrj.br}, 
M.A.B. do Vale\thanks{E-mail: aline@if.ufrj.br}\\
Instituto de F\'\i sica\\
Universidade Federal do Rio de Janeiro, RJ, Brazil \\}

\maketitle
\begin{abstract}
\par
The discovery potential for detecting new heavy Majorana and Dirac neutrinos at some recently proposed high energy $e^+e^-$ colliders is discussed. These new particles are suggested by grand unified theories and superstring-inspired models. For these models the production of a single 
heavy neutrino is shown to be more relevant than pair production when comparing cross sections 
and neutrino mass ranges. The process $ e^+e^- \longrightarrow {\nu} \, e^{\pm}\, W^{\mp}$ is calculated including on-shell and off-shell heavy neutrino effects. We present a detailed study 
of cross sections and distributions that shows a clear separation between the signal and standard model contributions, even after including hadronization effects.
\end{abstract}
\vskip 1cm
\noindent{PACS: 12.60.-i, 13.85.-t, 14.60.-z}
\par

\begin{multicols}{2}

\par
The recent Super-Kamiokande results  \cite{SUK} provided a strong evidence for neutrino oscillations and non zero neutrino masses. This has motivated many theoretical models that imply new heavy neutrino states. These new particles are present in several grand unified extentions of the standard model such as $SO(10)$ or $E_6$. Many other new states are present in these models but neutrinos are expected to play a fundamental role in any consistent extended model. This comes mainly from the possibility that light neutrino masses can be connected to the Fermi and grand unified scales through $ m_{\nu}=v_{Fermi}^2/v_{GUT}$. This relation can be obtained from the 
"see-saw" mechanism with new heavy neutrino states. A fundamental point to be experimentally clarified is the Dirac or Majorana nature of neutrinos.
\par
The new heavy neutrino masses are experimentally bounded to be greater than 80-100 GeV \cite{ZUB,PDG} and the mixing with light neutrinos is expected to be small, even if there is some model dependence on these results. This means that these new neutral leptons could be detected only at the 
next generation of high energy colliders NLC at SLAC and TESLA at DESY. In this paper we turn our attention to these new possible lepton colliders \cite{JMS}. New linear $e^+e^-$ high energy colliders have been proposed, with  center of mass energy from 500 GeV up to a few TeV. More recently $\mu^+\mu^-$ and $e^- e^-$ options were also proposed, as well as the electron-muon colliders.

\par
Some time ago it was noticed \cite{ARS,DJO} that for some models the single heavy lepton 
production in $e^+ e^- \longrightarrow {\nu} N$ is higher than pair production 
$e^+ e^- \longrightarrow N N$. Two main factors contribute for this difference. The first one is the mixing angle ($s_i \equiv \sin \theta_{mix}$) single power in the light-to-heavy neutrino vertex, contrary to the double mixing angle power in the heavy-to-heavy neutrino vertex. The second factor is phase space suppression. If we suppose that all mixing angles are of the same order, then we have in Table I the vertices for heavy neutrino interactions in three different models \cite{ARS} for new heavy neutrino states: vector singlet (VSM), vector doublet (VDM) and fermion-mirror-fermion models (FMFM). 
We call attention to the suppression factor for the $ZNN$ vertex in the vector singlet model, which is not present for the other models. 
Throughout this paper we will suppose that mixing 
angles for heavy-to-light neutrinos and new heavy neutrino masses are independent parameters. 
In the naive see-saw model, the mixing between light and heavy neutrinos is
given by $\displaystyle{\theta \simeq {m_{\nu}\over{ m_N}}}$. However, there are many
theoretical models which decouple the mixing from the mass relation \cite {BUC}. 
The mechanism is very simple. In the general mass matrix including Dirac and Majorana fields one imposes some internal symmetry that makes the matrix singular. Then the mixing
parameter has an arbitrary value, bounded only by its phenomenological consequences.
\par
Recently we have investigated this possibility for a single heavy Majorana
production in hadron-hadron colliders \cite{YSP}. It was shown that this mechanism is
more important than pair production and that like-sign dileptons can give a clear 
signature for this process, even after hadronization. The mixing of the presently known fermions and possible new heavy states is known to be small, of the order of $\sin^2 \theta_{mix}=10^{-2}-10^{-3}$. This comes from low energy phenomenology and the high precision measurements of the $Z$ properties at LEP/SLC. A recent estimate \cite{YPS} gives $\sin^2 \theta_{mix} < 0.0052$ with $95\%$ C.L. This limit value is used throughout this paper for all curves and distributions.

In this paper we present the complete first order treatment of single heavy neutral lepton production in electron-positron high energy colliders, including finite width effects. The full first order standard model background is also considered. We have studied the general process $e^+e^- \longrightarrow \nu_{\ell}\, \ell \, W $ where $\ell$ is a charged lepton and $\nu_{\ell}$ is a light neutrino or antineutrino. For the signal we can have the diagram contributions as shown in Fig. 1. Some earlier studies \cite{ARS,DJO,DNR} were done in the heavy neutrino on-shell approximation which includes diagrams (a) and (d) in Fig. 1. It is  known that for energies well above the Z mass the s-channel is suppressed and t-channel is dominant for N exchange. As we are interested in the study of distributions and experimental cuts, we have taken into account all the diagrams shown in Fig. 1 for the process $ 2 \longrightarrow 3 $. We can have lepton number conservation or violation, depending on the Dirac or Majorana nature of neutrinos, respectively. In this paper we consider  three different heavy neutrinos, one for each family. For Dirac heavy neutrinos we have lepton number conservation. Majorana neutrinos carry no internal quantum numbers and we are  then supposing  that all the vertices $NeW$ ; $N{\mu}W$ and $N{\tau}W$ have the same strength. In this case we are considering only the lighter Majorana  state. At the Fermi scale these new states are expected to behave as $SU_L(2)\otimes U_Y(1)$ basic representations.
\par
We can resume these interactions in the neutral and charged current lagrangians:

\begin{equation}
{\cal L}_{nc}=-\frac{g}{2c_W }Z_{\mu}\overline{\psi_i}\gamma^{\mu}
\left(g_V^{ij}-g_A^{ij}\gamma_{5}\right)\psi_j.
\end{equation}

and

\begin{equation}
{\cal L}_{cc}=-\frac{g}{2\sqrt{2}}
W_{\mu}\overline{\psi_i}\gamma^{\mu}\left(a^{ij}-b^{ij}\gamma_5\right)\psi_j.
\end{equation}

\bigskip
\noindent 
where i, j are the appropriate  combination of e, $\nu$, N with ``N'' the new neutral lepton. The light neutrinos couplings to the neutral Z are given by $ g_{V,A}=g_{V,A}^{SM} - {sin^2{\theta_{mix}}/2}$.
\par
The  decay modes for these leptons, in the Dirac case, are $ N_e \longrightarrow e^-\, W^+ $ and
$ N_e \longrightarrow \nu_e\, Z$.  For Majorana neutrinos \cite{YSP} we must include both signatures $N \longrightarrow \ell^{\mp}\, W^{\pm}$  and $ N \longrightarrow \nu_{\ell}\,(\bar\nu_{\ell})\, Z$, with $\ell=e,\mu,\tau$.

In Fig. 2 we show the total cross sections for single $e^+ e^- \longrightarrow {\nu} N $ and pair heavy lepton production $e^+ e^- \longrightarrow N N$ at $\sqrt s=$ 500 GeV. All curves are for on-shell heavy neutrinos. A single heavy Majorana has the higher cross section due to the sum over final neutrino and anti-neutrino production and to the sum over the three lepton families. The single heavy Dirac neutrino for the electron family dominates the associated muon (and tau) family production since in the first case we have s and t channel exchanges, whereas in the last cases we have only s channel contribution. Similar arguments apply to the pair production of heavy neutrinos. In the vector singlet model we have a strong suppression factor from the $ Z NN $ vertex in the $s$ channel. For the other models we also have an energy suppression in the amplitude from the Z propagator in the $s$ channel and a $ \sin^2 {\theta_{mix}}$ for the $t$ channel. In all cases, pair production is kinematically bounded to masses up to ${\sqrt s}/2$ whereas single heavy neutrino production can reach masses up to $ \sqrt s$.  For center of mass energy of 1 TeV a similar pattern of heavy neutrino production is present. For the more specific sub-process $e^+e^- \longrightarrow {\nu}\, e^+\, W^-$ a careful distinction between Dirac and Majorana neutrinos must be done and the correct set of Feynman diagrams must be chosen from Fig. 1. An important point on the Dirac or Majorana nature of the new possible heavy neutrino comes from Fig. 2. For a heavy Dirac neutrino the electron final state is two orders of magnitude greater than the final state muon.  For a Majorana heavy neutrino we expect an equal number of electrons and muons in the final state. Another point to be taken into account is the fact that the final state light neutrino is an experimentally undetected particle. So we must sum over all possible combinations whenever necessary. In all cases we have considered all contributions, with on-shell and off-shell single heavy neutrinos, as well as finite width effects, as recently investigated by Cvetic, Kim and Kim \cite{CVE}. The standard model background contributes with 12 diagrams, as shown in Fig. 3. Both the signal and background were calculated using the high energy program CompHep \cite{HEP}.
\par
Let us now turn our attention to the general characteristics of the process $ e^+e^- \longrightarrow {\nu}\, e^+\, W^-$. In Figs. 4-12 we have done simple detector cuts
$E_{lepton}> 5$ GeV and $-0.995 < cos{\theta_i} < 0.995$, where $\theta_i$ is the angle between
any final state particle and the initial electron. We define  $\theta_{e^+}$ as the angle between the initial electron and the final state positron; $\theta_{W^-}$ as the angle between the initial electron and the final state $W^-$ or hadrons and $\theta_{e^+ W^-}$ as the angle between the final positron and the final state $W^-$ or hadrons, in the $e^+ e^-$ center of mass frame.
 
In Fig. 4 we show the total cross section for the signal and standard model background, which is typically 1-1.5 orders of magnitude greater than the signal. One of the main points of our work is to show that this relation will be inverted by appropriated cuts in the $e^+ W^-$ invariant mass distribution. In order to justify our cuts, let us first look at the following angular distributions. In Figs. 5 and 6 we display the final state positron angular distribution (relative to the initial electron) for $M_N=$ 100 and $M_N=$ 400 GeV in the Majorana and Dirac cases, as well as the standard model contribution. We have taken these two mass values, one bellow and the other above
the pair production mass limit. We notice that for smaller masses, around 100 GeV, the Majorana and Dirac cases are different, but for higher masses this difference disappears. A similar situation is seen in Figs. 7 and 8 for the W angular distribution. The intermediate masses of heavy neutrinos can be easily inferred from Figs. 5-8.
\par
In Fig. 9 we show the angular distribution between the final $e^+$ and $W^-$ in the $e^+ e^-$ center of mass frame, for a heavy Majorana neutrino. 
The signal presents a clear kinematical bound from the Lorentz boost along the heavy neutrino direction. In the Dirac neutrino case we have a similar shape, above the Majorana curves.
\par
In order to have a more realistic estimate of the signal-to-background separation we have performed the $W^-$ hadronization using the Pythia program \cite{PIT}, for the signal and the standard model background. Another possible background source is the process $e^+ \, e^- \longrightarrow e^+ \,
e^- \, Z$ where the final state electron escapes detection. With the hadron jets peaked around the Z mass, we have verified that this channel has a small contribution to the dominant background. 
\par
The invariant visible mass ($e^+ \, +$ hadrons) 
versus missing (neutrinos) energy correlations are shown in Figs. 10-12, in arbitrary units. In Figs. 10-11-12, 
labeled ``a'' we have applied the following general detector cuts: $-0.95 < cos \theta_{e^+} < 0.995$ and $-0.995 < cos \theta_{W^-} < 0.95$. Here we have a very clear limited kinematical region coming from energy-momentum conservation. The signal is already present in these figures. The angular distributions in Figs. 5-9 suggests that the signal-to-background can be increased by applying more restrictive angular cuts. The results are shown in Figs. 10b.; 11b.; 12b. For a heavy neutrino mass of 100 GeV we have done $ -0.5 < cos  \theta_{e^+W^-} < 0.995 $ and the result is shown in Fig. 11a. For a mass $M_N=$ 200 GeV the results are shown in Fig. 11b., with angular cuts $ -0.9 < cos  \theta_{e^+} < 0.995 $; $ -0.5 < cos  \theta_{e^+W^-} < 0.5 $ ; $ -0.95 < cos  \theta_{W^-} < 0.9 $. For a heavy neutrino mass of 400 GeV we have Fig. 12b., with angular cuts $-0.5 < cos \theta_{e^+} < 0.995$; $-0.95 < cos \theta_{W^-} < 0.5 $; $ -0.95 < cos  \theta_{e^+W^-} < -0.7 $.  
These more restrictive angular cuts show quite clearly that the ratio signal/background can be improved. For center of mass energies of 1 TeV  we found analogous conclusions. 
\par
In conclusion, the present work shows that $e^+e^-$ colliders can test the existence of heavy Dirac and Majorana neutrino masses up to $\sqrt s$ in the $\nu \, e^{\pm}$ hadrons channel. For higher center of mass energies the conclusions are similar to the case of a $\sqrt s=$ 500 GeV presented in detail  in this paper, since the process is dominated by $t$ channel exchange. For $\mu^+ \mu^-$ colliders we have the same situation, if we replace the final $e^+$  by a $\mu^+$. Single heavy neutrino production bellow and above pair mass threshold can be clearly separated from the standard model background, even after hadronization and detector cuts. Angular cuts on the final state particle distributions can be applied unambiguously, in order to increase the signal-to-background ratio. As our estimate for the signal cross section uses the upper bound for $ \sin^2 {\theta_{mix}}$, an eventually negative experimental search can be converted into new more restrictive light-to-heavy neutrino mixing bound.

\vskip 2cm
{\it Acknowledgments:} This work was partially supported by the
following Brazilian agencies: CNPq, FUJB, FAPERJ and FINEP.

\vfill\eject

\vskip 1cm
\vspace{1cm}
\LARGE
Figure Captions
\normalsize
\begin{enumerate}
\item Signal Feynman graphs for heavy Majorana (N) and Dirac ($N_e ,N_\mu$) neutrino contribution to $e^+e^- \longrightarrow {\nu} \, e^+ \, W^-$.
\item Single and pair production of on-shell heavy Dirac (VSM,VDM and FMFM) and Majorana neutrinos at $\sqrt s=$ 500 GeV for $e^+e^-$ colliders ($\sin^2 \theta_{mix}=0.0052$).
\item Standard model background contribution to $e^+e^- \longrightarrow {\nu} \, e^+ \, W^-$.
\item Signal and background (standard model) contributions to $e^+e^- \longrightarrow {\nu} \, e^+\, W^-$ at $\sqrt s=$ 500 GeV.
\item Final positron angular distribution relative to the initial electron for heavy Majorana neutrinos with $M_N=$ 100 GeV and $M_N=$ 400 GeV.
\item Final positron angular distribution relative to the initial electron for heavy Dirac neutrinos with $ M_N=$ 100 and $M_N=$ 400 GeV.
\item $W^- $ angular distribution relative to the initial electron for heavy Majorana neutrinos with $M_N=$ 100 GeV and $M_N=$ 400 GeV.
\item $W^-$ angular distribution relative to the initial electron for heavy Dirac neutrinos  
with $M_N=$ 100 GeV and $M_N=$ 400 GeV.
\item Angular distribution between final positron and $W^-$ for heavy Majorana neutrinos with  
$M_N=$ 100 GeV, $M_N=$ 200 GeV and $M_N=$ 400 GeV.
\item Invariant visible mass ($e^+ \, +$ hadrons) versus missing (neutrino) energy correlation for background and signal 
for $M_N=$ 100 GeV (in arbitrary units). Fig. 10a. was done with the general cuts and Fig. 10b. was done with the additional cuts as discussed in the text.  
\item Same as figure 10 for $M_N=$ 200 GeV.
\item Same as figure 10 for $M_N=$ 400 GeV.

\end{enumerate}
\end{multicols}

\begin{table}
\begin{center}
\begin{tabular}{|c|l|l|l|} 
  & VSM  & VDM  & FMFM \\  \hline
$W\longrightarrow e N$ & $a=s_i$ &  $a=s_i$ & $a=0$\\
 & $b=s_i$ & $b=-s_i$ & $b=2s_i$ \\ \hline
$Z\longrightarrow \nu N$ & $g_V=s_i/2$ & $g_V=s_i/2$ & $g_V=0$ \\
 & $g_A=s_i/2$ & $g_A=-s_i/2$ & $g_A=s_i$ \\ \hline
$Z\longrightarrow N N$ & $g_V=s^2_i/2$ & $g_V=1$ &  $g_V=1/2$  \\
 & $g_A=s^2_i/2$ & $g_A=s^2_i/2$ & $g_A=-1/2$ \\ 
\end{tabular}
\caption{ Mixing angles for $W$ and $Z$ couplings with new neutral heavy leptons for three different models.}
\end{center}
\end{table}
\end{document}